\documentclass{article}

\usepackage{amsfonts}
\usepackage{amsmath}
\usepackage{amsthm}
\usepackage{amssymb}

\def\be{\begin{eqnarray}}
\def\ee{\end{eqnarray}}
\def\nn{\nonumber}

\newcommand{\ZG}{\mathbb{Z}}

\newcommand{\M}{\mathcal{M}}
\newcommand{\RF}{\mathbb{R}}
\newcommand{\bd}[1]{\left\{#1\right\}}
\newcommand{\br}[1]{\left(#1\right)}

\newcommand{\su}[1]{\mathop{\sum}_{#1}}
\newcommand{\summ}[2]{\displaystyle \mathop{\sum}_{#1}^{#2}}
\newcommand{\prodd}[2]{\displaystyle \mathop{\prod}_{#1}^{#2}}

\newcommand{\al}{\alpha}
\newcommand{\g}{\gamma}
\newcommand{\genu}[1]{^{(#1)}}
\newcommand{\Hs}{\mathcal{H}}

\newcommand{\eq}[1]{\begin{equation} #1 \end{equation}}
\newcommand{\eqm}[1]{\begin{multline} #1 \end{multline}}

\newcommand{\eql}[2]{\begin{equation} \label{e:#1} #2 \end{equation}}

\newcommand{\eqs}[1]{\begin{align} #1 \end{align}}

\newcommand{\en}[1]{\begin{enumerate}  #1 \end{enumerate}}

\newcommand{\re}[1]{(\ref{e:#1})}

\newcommand{\ths}{\vartheta}
\newcommand{\defeq}{\displaystyle \mathop{=}^{def}}
\newcommand{\ep}{\varepsilon}
\newcommand{\de}{\delta}
\newcommand{\ch}[2]{\left[\begin{array}{c}#1\\ #2 \end{array}\right]}

\begin{document}

\hfill ITEP/TH-35/09

\bigskip

\centerline{\Large{
Lattice Theta Constants vs Riemann Theta Constants
}}
\centerline{\Large{and NSR Superstring Measures}}

\bigskip

\centerline{P.Dunin-Barkowski\footnote{E-mail: barkovs@itep.ru}, 
A.Morozov\footnote{E-mail: morozov@itep.ru} and 
A.Sleptsov\footnote{E-mail: sleptsov@itep.ru}}

\bigskip

\centerline{\it ITEP and MIPT, Moscow and Dolgoprudny, Russia}

\bigskip

\centerline{ABSTRACT}

\bigskip

{\footnotesize
We discuss relations between two different representations
of hypothetical holomorphic NSR measures,
based on two different ways of constructing
the semi-modular forms of weight 8.
One of these ways is to build forms from
the ordinary Riemann theta constants
and another -- from the lattice theta constants.
We discuss unexpectedly elegant relations between lattice
theta constants, corresponding to 16-dimensional self-dual lattices,
and Riemann theta constants
and present explicit formulae expressing the former ones
through the latter.
Starting from genus 5 the modular-form approach
to construction of NSR measures runs into serious problems
and there is a risk that it fails completely already at genus 6.
}

\bigskip

\bigskip

\tableofcontents

\section{Introduction}

After the famous work of A.Belavin and V.Knizhnik \cite{BK}
it turned out that perturbative string theory
can by built using the Mumford measure \cite{MumM} on the moduli space
of algebraic curves,
that is summation over all world-sheets can be replaced
by integration over the moduli space with this measure.
For NSR superstring there should be
a whole collection of measures for every genus \cite{superst},
corresponding to different boundary conditions for fermionic
fields.
They are labeled by semi-integer theta-characteristic:
collections of zeroes and unities, associated with
non-contractable cycles of the Riemann surface.
While bosonic string could be studied
without Belavin-Knizhnik theorem, for super- and heterotic strings
this is absolutely impossible, because GSO projection
requires to sum holomorphic NSR measures over characteristics
before taking their bilinear combinations with complex conjugate
measures.

It is a long-standing problem to find these NSR superstring measures
(see \cite{Mrev1} for a recent review and numerous references).
The direct way to derive them from the first principles
turned to be very hard,
and it took around 20 years before E.D'Hoker and D.Phong
in a long series of papers managed to do this
in the case of genus 2 \cite{DHP}.
Higher genera seem to be even harder.
However, there is another, seemingly simpler approach to the problem:
to {\it guess} the answer from known physical and mathematical
requirements which it is supposed to satisfy.
This approach proved to be quite effective in the case of
bosonic strings \cite{bosM} and, after the new
insight from \cite{DHP} it was successfully applied to the
case of NSR measures in \cite{Ita}-\cite{OPSMY}.
In the present paper we discuss ans\"atze for NSR
measures, which were proposed following this second way
of attacking the problem.

In \cite{bosM} the physical problem {\it for low genera}
was reformulated as one in the theory of modular forms:
one needs to construct semi-modular forms of the weight $8$
with certain properties.
There are two obvious ways to represent such forms:
through {\it ordinary} Riemann theta-constants and through
{\it lattice} theta-constants.
Accordingly there are two ans\"atze for NSR measures at low genera.
One -- in terms of Riemann theta constants --
was suggested in \cite{DHP,Ita} and in its final appealing
form by S. Grushevsky in \cite{Gru}
and the other -- in terms of lattice theta constants,
associated with the eight
(six odd and two even) 16-dimensional unimodular lattices. --
by M.Oura, C.Poor, R.Salvati Manni and D.Yuen in \cite{OPSMY}.

In fact the theory of modular forms is hard, it is being built
anew for every new genus, and it is even
a question which benefits more from the other:
string theory from the theory of modular forms or vice versa.
In this paper we continue the study
of relations between
lattice theta constants and Riemann theta constants.
We write explicit formulae expressing the lattice theta constants
through Riemann theta constants in all genera and
for all odd lattices except one.
For this remaining one we write explicit formulae
in genera $g \leq 4$ and discuss a hypothesis of how
it may look in higher genera. These formulae look as follows:
\eq{
\ths_p^{(g)} = 2^{-gp} \xi^{(g)}_p, \quad p=0..4,
}
\eq{
\ths_5^{(g)} = 2^{-\frac{g(g-1)}{2}}\cdot
\br{\prodd{i=1}{g}\br{2^i-1}}^{-1} G^{(g)}_g
}
(see s.\ref{s:grush} for description of our notation).
It is interesting that for the first five odd lattices the proof
is provided by a simple rotation of the lattice with the help
of the so-called Hadamard matrices.

In other parts of the paper we discuss modular-form
ans\"atze for NSR measures and relations between them.
According to \cite{BK,bosM} the measures are written as
\eq{
d\mu[e] = \Xi[e] d\mu,
\label{mfa}}
where $d\mu$ is the Mumford measure,
and they are subjected to the following conditions:
\en{
\item $\Xi[e]$ are (semi-)modular forms of weight 8,
at least for low genera,
\item they satisfy factorization property when Riemann surface degenerates,
\item the "cosmological constant" should vanish: $\sum_e \Xi[e] = 0$,
\item for genus $1$ the well known answer should be reproduced.
}
For detailed description of these properties see the next section.

Let us briefly describe the modern ans\"atze
\cite{Ita}-\cite{OPSMY}, inspired by \cite{DHP}.

In Grushevsky ansatz $\Xi[e]$ is written
as a linear combination of functions $\xi_p[e]$,
which are sums over sets of $p$ characteristics of monomials
in Jacobi theta constants of order 16.
This is literally true for genera $g \leq 4$,
for $g > 4$ roots of monomials in Riemann theta constants
appear, but the total degree remains $16$.
This ansatz may in principle be written for any genus,
however it satisfies all conditions only for genera $g \leq 4$.
At genus 5 in its pure form it fails to satisfy the cosmological
constant property \cite{GruSM} (see also \cite{GabVo}).

In OPSMY ansatz expressions for $\Xi[0]$ are written
in terms of lattice theta constants, corresponding to six
odd unimodular lattices of dimension 16.
OPSMY showed that for $g \leq 4$ these lattice theta constants
span the entire space of holomorphic functions on the Siegel half-space
$\mathcal{H}$, which are modular forms under
$\Gamma_g\br{1,2}$.
However, because the number of these lattices is fixed,
they doubtly span this space for $g \geq 6$,
because the dimension of this space can grow (in fact, it is known
that it doesn't grow starting from $g \geq 17$; in principle it is
possible that this dimension becomes stable already at genus 5).
This fact that the number of lattice theta constants is limited
leads to the following: OPSMY ansatz cannot be written for $g \geq 6$
because in these cases one cannot compose from lattice theta constants
anything that would satisfy factorization constraint.

Both Grushevsky and OPSMY ans\"atse potentially
have problems in genus 5, because it turned out that
in their pure form
the cosmological constant does not vanish.
Grushevsky and OPSMY then propose to resolve this problem
by brute force: if we have a sum of several terms,
which should be zero, but apparently is not,
\eq{
\su{e}\Xi[e]^{(5)} \br{\tau} = F^{(5)}\br{\tau} \neq 0,
}
then we can subtract from each term the sum,
divided by number of terms.
Then the sum of the modified terms will obviously vanish:
\eq{
\su{e}\widetilde{\Xi}[e]^{(5)} \br{\tau}
= \su{e}\br{\Xi[e]^{(5)} \br{\tau} - \frac{1}{N_{even}}F^{(5)}\br{\tau}}= 0
}
This indeed solves the problem in genus 5,
because in this particular case such change of
$\Xi[e]$ does not spoil the factorization constraint,
since $F$ for all genera $g<5$ vanishes on moduli subspace in the
Siegel half-space and thus vanishes whenever the genus-5
surface degenerates.
Thus at genus 5 there can be a way out of cosmological
constant problem, at least formally.

One can speculate that this solution for genus 5 doesn't seem
very natural. Note that following considerations about it
are only intuitive arguments and by no means have status of
mathematical theorems.
For all genera $g \leq 3$ the moduli space
(i.e. the Jacobian locus) coincides with the Siegel half-space.
Therefore the cosmological constant, being zero on moduli,
vanishes on entire Siegel space.
However, for genus $4$ the moduli space becomes non-trivial:
it is a rather sophisticated subspace of codimension one
in the Siegel half-space.
And in this case cosmological constant vanishes only on moduli,
but remains nonzero at other points of the Siegel half-space.
Note that since for NSR measures values of $\Xi[e]$ are
important only on moduli, one can easily add to each term a
form which is vanishing on Jacobian locus, for example, minus
the cosmological constant divided by the number of terms.
That is while the solution for NSR measures problem is unique,
its representation as Siegel forms is not unique at all.
Thus, strictly speaking, we cannot say anything about
cosmological constant outside moduli, because it depends
on this choice of continuation of functions $\Xi[e]$.
What we mean by saying that cosmological constant
``remains nonzero at other points of the Siegel half-space''
is that it equally remains nonzero for both ans\"atze for NSR measures
if we think of the forms $Xi[e]$ provided by them as forms
on the Siegel space. That is, if one just tries to write
an ansatz for NSR measures through Siegel modular forms
satisfying all conditions, one in $g=4$ case will naturally
obtain the Schottky form as cosmological constant.
It may be natural to expect the same for $g=5$:
that cosmological constant also does not vanish on
the entire Siegel half-space in genus 5.
However the ansatz, obtained by subtraction technique
makes cosmological constant zero on \textit{entire}
$\mathcal{H}$ (note here that both Grushevshsky and OPSMY
prove that the expression for subtraction term they use is
equal to the sum of $\Xi[e]$ only on Torelli space, which
is a covering of the moduli space, and not on entire Siegel
half-space; thus our last statement is somewhat ungrounded).
This behavior makes this solution little bit unnatural.
This feeling is confirmed by the fact that this method fails
to work for genus 6 and higher:
if we try to resolve the problem with cosmological constant
in the same way,  we lose the factorization property
already at $g=6$.

It deserves emphasizing that there is {\it no} reason to
believe that $\Xi[e]$ in (\ref{mfa}) is expressed through
modular forms for $g>4$. One could only hope and try.
As we see, the result seems negative, still a lot of
interesting details can be learnt in the process.
Anyhow, at this moment it is unclear
what to do with the superstring measures at genus 6 and above
(and the suggested answer for $g=5$ is also not very
convincing).

The structure of the paper is as follows.
In Section \ref{s:problem} we discuss mathematical conditions
on NSR superstring measures.
In Section \ref{s:grush} Riemann theta constants are
introduced and Grushevsky ansatz is described.
In the next section \ref{s:opsmy} lattice theta constants, associated with
self-dual 16-dimensional lattices, are introduced and OPSMY ansatz is discussed.
Section \ref{s:ltcvsotc} is central to the paper
and is devoted to explicit relation between lattice and
ordinary theta constants.
The following Section \ref{s:sl} is about the strange behaviour
of the theta constant, associated with the sixth odd lattice
$\br{D_8\oplus D_8}^+$. And finally in the last Section \ref{s:arel}
we discuss the relation between Grushevsky and OPSMY ans\"atze.
We show {\it explicitly} that Grushevsky and OPSMY
ans\"atze coincide for genera $g \leq 4$,
which is in perfect agreement with the uniqueness properties
proved by both Grushevsky and OPSMY.
Then we discuss the relation between the ans\"atze for genus 5.

\section{The problem of finding superstring measures}
\label{s:problem}

In this section we discuss the mathematical problem of finding the
NSR superstring measures in the framework of the modular-form hypothesis
\cite{bosM}.

First of all, we review the setting.
Superstring measures at genus $g$, which we are intended to find,
form a set of measures $\bd{d\mu_e}$ on the moduli space
of algebraic curves of genus $g$.
Here index $e$ stands for an \textit{even characteristic},
labeling boundary conditions for fermionic fields on the curve
\cite{superst}.
A \textit{characteristic} is by definition
a collection of two $g$-dimensional $\ZG_2$-vectors,
i.e. $e \in \br{\ZG_2^g}^2$. We write
\eq{
e = \ch{\vec\de}{\vec\ep},
}
where $\vec\de,\vec\ep \in \ZG_2^g$.
A characteristic is called \textit{even}
if the scalar product $\vec\delta\cdot\vec\ep$ is even.
There are $N_{even}=2^{g-1}\br{2^g+1}$ even characteristics in genus $g$.

On the moduli space there is already a distinguished measure
-- the \textit{Mumford measure} $d\mu$ \cite{BK,MumM,bosM}.

It was proposed in \cite{BK,bosM} that superstring measures are expressed
in terms of the Mumford measure in the following way:
\eq{
d\mu_e = \Xi_e d\mu,
}
with $\Xi_e$ being some functions on moduli space,
satisfying certain conditions.

To pass to this conditions we shall first introduce a convenient way
to describe {\it some} functions on moduli space.
Period matrices of the curves, corresponding to particular points
in the moduli space, lie in the so-called \textit{Siegel half-space}
$\Hs\genu{g}$. $\Hs\genu{g}$ is simply the space of $g\times g$ symmetric
complex matrices $\tau$ with positive definite imaginary part
and has (complex) dimension $\frac{g(g+1)}{2}$.
Period matrices form the so-called \textit{Jacobian locus}
$\M$ inside it, which is a submanifold
(strictly speaking, sub\textit{orbi}fold)
of complex dimension $3g-3$
(for $g \geq 2$; for $g=1$ the dimension is also 1).
We then can identify the moduli space with the Jacobian locus
and consider a class of functions on moduli space,
which depend on $\tau$, i.e. which can be somehow analytically
continued to $\Hs\genu{g}$.
One should be cautious here:
not all the functions on moduli space are of this type,
most important, the ones that arise in the free-field calculus
on complex curves \cite{frec}, and are the building blocks
for the string measures in straightforward approach,
do not usually belong to this class. Still, at least at
low genera, they combine nicely into functions, which
depend only on $\tau$
(the first {\it non-trivial} example of this kind was the formula
for the Mumford measure at genus 4 in \cite{bosM} -- actually not
proved in any alternative way till these days),
and this motivated the search for $\Xi[e]$ inside this class
\cite{oldpap} and nearby \cite{oldpapvic}.

Also let us define the notion of \textit{modular form},
since under above hypothesis $\Xi[e]$ would be of such type.
The modular group is defined as $\Gamma\genu{g} \defeq \mathrm{Sp}(2g,\ZG)$
and it acts on $\Hs\genu{g}$ as follows:
\eq{
\g = \left(\begin{array}{cc}
A & B \\
C & D
\end{array} \right) \in \Gamma\genu{g},
}
\eq{
\g: \tau \mapsto \br{A\tau + B}\br{C\tau+D}^{-1}.
}
Then holomorphic function $f$ on $\Hs$ is called a
\textit{modular form of weight $k$ with respect to subgroup $\Gamma'$
of the modular group} if it satisfies the following property:
\eq{
f\br{\g\tau} = \det\br{C\tau+D}^{k} f\br{\tau}
}
for all $\g \in \Gamma'$.

We will be interested in one particular subgroup of the modular group, 
which is called $\Gamma(1,2)$. It is defined as follows: an element
\eq{
\g = \left(\begin{array}{cc}
A & B \\
C & D
\end{array} \right)
}
of the modular group belongs to $\Gamma(1,2)$ iff all elements 
on diagonals of matrices $AB^T$ and $CD^T$ are even. This subgroup 
is interesting in conjunction with theta constants, which will be 
defined in following sections. Now we can say that while action of 
a general element of the modular group on a Riemann theta constant 
changes its characteristic, the action of an element of 
$\Gamma(1,2)$ leaves zero characteristic invariant.

We finally come to the announced conditions on functions $\Xi_e$
and are eventually able to formulate the problem about
superstring measures as a well-posed mathematical problem.

So, the question is as follows:
are there any functions $\Xi_e$ in every genus $g$,
which satisfy the following four properties?
\en{
\item $\Xi_e$ is a modular form of weight $8$
with respect to subgroup $\Gamma_e \subset \Gamma\genu{g}$,
at least when restricted to the Jacobian locus.
Sometime such forms are called semi-modular.
The subgroup is defined as
$\Gamma_e \defeq \gamma[e] \Gamma_g\br{1,2} \gamma[e]^{-1}$, 
where $\gamma[e]$ is an element of $\Gamma_g$,
which transforms the zero characteristic to characteristic $e$.
Saying that an element of the modular group acts on a characteristic,
we mean that it acts on the \textit{Riemann theta constant},
associated with this characteristic and transforms it into
Riemann theta constant with another characteristic.
Riemann theta constants will be defined in Section \ref{s:grush}.
\item $\Xi_e$ satisfies the following \textit{factorization property}:
\eq{
\Xi^{(g)}[e]\left(\begin{array}{cc} \tau^{(g_1)} & 0 \\
0 & \tau^{(g-g_1)}\end{array}\right) =
\Xi_{e_1}^{(g_1)}\br{\tau^{(g_1)}}\,\Xi_{e/e_1}^{(g-g_1)}\br{\tau^{(g-g_1)}}
}
\item $\Xi[e]$ satisfies the
\textit{property of vanishing cosmological constant}:
\eq{
\su{e}\Xi_e\br{\tau} = 0
\label{cococo}}
One should keep in mind that
the naming convention for this property is a bit abused,
because cosmological constant is actually the integral
of the total measure over all string configurations.
Here it is \textit{not} the case: when we say that
cosmological constant vanishes, we mean that this happens point-wise:
the total measure is zero at \textit{every point}
$\tau$ of the moduli space.
\item
For genus one $\Xi_e$ reproduces the known answer from elementary
superstring theory \cite{superst}:
\eq{
\Xi_e\genu{1} = \theta^4[e] \prod_{e'}^3 \theta[e']^4 =
\theta^{16}[e] - \frac{1}{2}\theta^8[e]\br{\su{e'}\theta^8[e']}
}
Here $\theta[e]\br{\tau}$ stands for Jacobi theta constant,
see Section \ref{s:grush} below.
This property is important, because factorization condition
iteratively reduces all measures to genus one.
}

Addressing this mathematical problem is actually an attempt
to guess the answer for superstring measures from their known properties
instead of doing the calculation from the first principles,
which is very difficult in higher genera.
We would face a difficulty on this way if it turned out that there are
several different possible collections $\Xi_e$ in some genus,
satisfying all these conditions.
However, this is not the case, moreover the situation is quite
the opposite.
There are two ans\"atze proposed, which give the same results up
to the genus 5 and do not work further.
Thus today it looks like it is impossible to give the answer
for genera $g\geq 6$ in such terms.
The answer to the overoptimistic question,
if superstring measures or, more carefully, the
ratios $d\mu[e]/d\mu$, can be represented as
semi-modular forms on entire Siegel half-space
for all genera, now seems to be negative.
We should look either for more sophisticated combinations
of modular forms, including residues of Schottky forms like
in bosonic measure at genus four, or even switch to the
functions, which are modular forms only when restricted to Jacobian locus.

In the two following sections we review the two available ans\"atze
and then use them to illustrate this negative claim.

\section{Grushevsky ansatz}
\label{s:grush}

First of all, we need to introduce the \textit{Riemann theta constants},
which are the functions on the Siegel half-space,
in terms of which the ansatz is written.
Note that this usual terminology is rather misleading here,
because they are indeed functions, not constants on the Siegel half-space.
This naming convention reflects the fact that there is a notion of
\textit{"Riemann theta function"} which is a function
not only of the modular parameter $\tau$, but also of coordinates $\vec z$
on the Jacobian (a $g$-dimensional torus),
associated to this value of modular parameter.
In this context the theta functions with $\vec z=0$
are usually called "theta constants",
because they do not depend on $\vec z$.
In the present paper we will use only theta constants
-- and call them {\it constants} to avoid possible confusion,
despite we need and use them as {\it functions} on the Siegel half-space.

So the Riemann theta constant with characteristic at genus $g$
is defined as follows \cite{MumB}:
\eq{
\theta\ch{\vec\de}{\vec\ep} \br{\tau} \defeq \su{\vec n \in \ZG^g}
\exp\left(\pi i \br{\vec n+\vec\delta/2}^T\tau\br{\vec n+\vec\delta/2} +\pi i
\br{\vec n+\vec\delta/2}^T\vec\ep\right)
}
All Riemann theta constants with odd characteristics are
identically zero.

It turns out that Riemann theta constants
are the nice building blocks for modular forms on the Siegel half-space.
More precisely, they behave in the following way under modular
transformations $\gamma \in \Gamma(1,2)$:
\eq{
\theta\ch{D\vec\de-C\vec\ep}{-B\vec\de+A\vec\ep}\br{\g\tau} = 
\zeta_{\gamma} \det\br{C\tau+D}^{1/2}
\exp\br{\frac{\pi i}{4}\br{2\vec\de^T B^TC\vec\ep-\vec\de^T 
B^TD\vec\de-\vec\ep^T A^TC\vec\ep}}
\theta\ch{\vec\de}{\vec\ep}\br{\tau},
}
where $\zeta_{\gamma}$ is some eighth root of unity which 
depends only on $\g$.
It is then straightforward to see that $\theta^{16}[0]$
is a modular form of weight 8 w.r.t. $\Gamma\br{1,2}$. The 
sum of $\theta^{16}[e]$ over all even characteristics
will be a modular form of weight 8 w.r.t. the whole modular group.
When we write $[0]$ we everywhere mean zero characteristic,
i.e. characteristic, for which all elements of both vectors
$\vec\delta$ and $\vec\ep$ are zeroes.

In any genus Grushevsky ansatz can be expressed
through the following combinations of Riemann theta constants
\cite{Ita, Mrev1}
(for brevity we write characteristics as indices in the r.h.s.)
\eqs{
\xi^{(g)}_0[e] &= \theta_e^{16}, \nn \\
\xi^{(g)}_1[e] &= \theta_e^8 \sum_{e_1}^{N_e} \theta^8_{e+e_1}, \nn \\
\xi^{(g)}_2[e] &= \theta_e^4\sum_{e_1,e_2}^{N_e}
\theta^4_{e+e_1}\theta^4_{e+e_2}\theta^4_{e+e_1+e_2}, \nn \\
\xi^{(g)}_3[e] &= \theta_e^2\sum_{e_1,e_2,e_3}^{N_e}
\theta^2_{e+e_1}\theta^2_{e+e_2}\theta^2_{e+e_3}
\theta^2_{e+e_1+e_2}\theta^2_{e+e_1+e_3}\theta^2_{e+e_2+e_3}
\theta^2_{e+e_1+e_2+e_3}, \nn \\
\ldots
\label{xi8p1}
}
\be
\xi^{(g)}_p[e] = \sum_{e_1,\ldots,e_p}^{N_e}\left\{\theta_e\cdot
\left(\prod_i^p \theta_{e+e_i}\right)\cdot
\left( \prod_{i<j}^p\theta_{e+e_i+e_j}\right)\cdot
\left(\prod_{i<j<k}^p \theta_{e+e_i+e_j+e_k}\right)\cdot \ldots
\cdot\theta_{e+e_1+\ldots+e_p}\right\}^{2^{4-p}}
\nn
\ee
in the following way:
\eq{
\Xi^{(g)}[0] = \summ{p=0}{g}h^{(g)}_p\xi_p^{(g)}[0],
}
where
\eql{gran}{
h_p^{(g)} = (-1)^p\cdot 2^{\frac{(g-p)^2 -(g+p)}{2}}\cdot
\br{\displaystyle \prodd{i=1}{p}\br{2^i-1}\prodd{i=1}{g-p}\br{2^i-1}}^{-1}
}
Other $\Xi[e]$ are obtained just by substituting $e$ instead of $0$
into $\xi[0]$.
Formula \re{gran} becomes more compact if
written in terms of the so-called 2-factorial numbers
(for definition of $q$-factorials and $q$-binomial coefficients see,
for example, \cite{Morrison}):
\eq{
h_p^{(g)} = \frac{(-1)^p
}{[p]_2![g-p]_2!}\cdot
2^{\frac{(g-p)^2 -(g+p)}{2}}
}
Explicitly the first five forms look like
(when we write $\xi_p$ without characteristic, we mean $\xi_p[0]$)
\eql{gra1}{
\Xi^{(1)}[0]=\xi_{{0}}-\frac{1}{2}\,\xi_{{1}},
}
\eq{
\Xi^{(2)}[0]=\frac{2}{3}\,\xi_{{0}}-\frac{1}{2}\,\xi_{{1}}+
\frac{1}{12}\,\xi_{{2}},
}
\eq{
\Xi^{(3)}[0]={\frac {8}{21}}\,\xi_{{0}}-\frac{1}{3}\,\xi_{{1}}+
\frac{1}{12}\,\xi_{{2}}-{\frac {1}{168}}\,\xi_{{3}},
}
\eql{gra4}{
\Xi^{(4)}[0]={\frac {64}{315}}\,\xi_{{0}}-{\frac {4}{21}}\,\xi_{{1}}+
\frac{1}{18}\,\xi_{{2}} -{\frac {1}{168}}\,\xi_{{3}}+{\frac {1}{5040}}\,\xi_{{4}},
}
\eql{gra5}{
\Xi^{(5)}[0]={\frac {1024}{9765}}\,\xi_{{0}}-{\frac {32}{315}}\,\xi_{{1}}+
{\frac {2}{63}}\,\xi_{{2}}-{\frac {1}{252}}\,\xi_{{3}}+
{\frac {1}{5040}}\,\xi_{{4}}-{\frac {1}{312480}}\,\xi_{{5}}
}

Alternatively one can use the so-called Grushevsky basis \cite{Gru,Mrev1}:
\eqs{
G^{(g)}_0[e] &= \theta_e^{16}, \nn \\
G^{(g)}_1[e] &= \theta_e^8 \sum_{e_1 \neq 0}^{N_e} \theta^8_{e+e_1},\nn\\
G^{(g)}_2[e] &= \theta_e^4\sum_{e_1\neq e_2 \neq 0}^{N_e}\theta^4_{e+e_1}
\theta^4_{e+e_2}\theta^4_{e+e_1+e_2}, \nn \\
G^{(g)}_3[e] &= \theta_e^2\sum_{e_1 \neq e_2 \neq e_3 \neq 0}^{N_e}
\theta^2_{e+e_1}\theta^2_{e+e_2}\theta^2_{e+e_3} \theta^2_{e+e_1+e_2}
\theta^2_{e+e_1+e_3}\theta^2_{e+e_2+e_3}\theta^2_{e+e_1+e_2+e_3}, \nn \\
\ldots
}
\be
G^{(g)}_p[e] = \sum_{e_1 \neq \ldots \neq e_p \neq 0}^{N_e}\left\{\theta_e\cdot
\left(\prod_i^p \theta_{e+e_i}\right)\cdot
\left( \prod_{i<j}^p\theta_{e+e_i+e_j}\right)\cdot
\left(\prod_{i<j<k}^p \theta_{e+e_i+e_j+e_k}\right)\cdot \ldots
\cdot\theta_{e+e_1+\ldots+e_p}\right\}^{2^{4-p}}
\nn
\ee
Here sums are taken over sets of characteristics in which all characteristics 
are different.
This basis is related to $\xi^{(g)}_p$ basis as follows:
\eqs{
G^{(g)}_p[e] &= \summ{k=0}{p}(-1)^{k+p}\cdot 2^{\frac{(p-k)(p-k-1)}{2}}
\cdot \frac{\prodd{i=1}{p}\br{2^i-1}}{\prodd{i=1}{k}\br{2^i-1}
\prodd{i=1}{p-k}\br{2^i-1}}\xi^{(g)}_k[e] =\nn \\
 &= \summ{k=0}{p}(-1)^{k+p}\cdot 2^{\frac{(p-k)(p-k-1)}{2}}
 \br{\begin{array}{c} p\\k \end{array}}_{\hspace{-3pt}2}\xi^{(g)}_k[e],
}
where in the last line the 2-binomial coefficients were used
(also known as Gaussian binomial coefficients for $q=2$).

Expressions for $\Xi^{(g)}[0]$ are then written in terms of Grushevsky basis as
\eq{
\Xi^{(g)}[0] = \summ{p=0}{g}f^{(g)}_p G^{(g)}_p[0],
}
where
\eqs{
f^{(g)}_p &= (-1)^p\cdot 2^{-g} \cdot \br{\prodd{i=1}{p}\br{2^i-1}}^{-1} =\\
 &= (-1)^p\cdot 2^{-g} \cdot \frac{1}{[p]_2!}.
}
Explicitly the first five lines look like
\eq{
\Xi^{(1)}[0] = \frac{1}{2}\br{G_{{0}}-G_{{1}}},
\nn}
\eq{
\Xi^{(2)}[0] = \frac{1}{4}\br{G_{{0}}-G_{{1}}+\frac{1}{3}\,G_{{2}}},
\nn}
\eq{
\Xi^{(3)}[0] = \frac{1}{8}\br{G_{{0}}-G_{{1}}+\frac{1}{3}\,G_{{2}}-
\frac{1}{21}\,G_{{3}}},
\nn}
\eq{
\Xi^{(4)}[0] = \frac{1}{16}\br{G_{{0}}-G_{{1}}+\frac{1}{3}\,G_{{2}}-
\frac{1}{21}\,G_{{3}}+{\frac {1}{315}}\,G_{{4}}},
\nn}
\eq{
\Xi^{(5)}[0] = \frac{1}{32}\br{G_{{0}}-G_{{1}}+\frac{1}{3}\,G_{{2}}-
\frac{1}{21}\,G_{{3}}+{\frac {1}{315}}\,G_{{4}}-{\frac {1}{9765}}\,G_{{5}}}
}
When we write $G_p$ without characteristic, we mean $G_p[0]$.

Let us denote the sum of all $\Xi[e]$ of the Grushevsky ansatz,
i.e. the proposed cosmological constant, as $F^{(g)}$ in genus $g$.
It can be proved \cite{GruSM} that it is equal (up to a constant factor)
to the following celebrated expression in theta constants \cite{BK,bosM}:
\eq{
F^{(g)} = 2^g\su{e}\theta^{16}[e]-\br{\su{e}\theta^8[e]}^2,
\label{Fdef}}

It can be checked that Grushevsky ansatz satisfies the above requirements
for superstring measures in genera $g \leq 4$.
However it fails to do so in genus 5 because the cosmological constant
$F^{(5)}$ turns to be nonzero \cite{GruSM}.
Grushevsky proposes a way to overcome this problem:
consider instead of $\Xi[e]$ functions $\Xi[e]-N_{even}^{-1} F$.
This obviously solves the cosmological constant problem and
moreover does not spoil other properties in the case of genus 5.
The factorization property is not spoilt because $F^{(g)}$
is zero on the Jacobian locus for all genera up to 4 includingly,
and thus is zero on points of the Jacobian locus of the Siegel space
of genus 5 that correspond to factorization.

So, with a bit of modification the ansatz apparently works
up to genus 5 includingly.
However for genus 6 and above the same problem persists
and can not be cured anymore in the above-mentioned subtraction way
even formally,
because it spoils the factorization property for that cases:
when a genus-six curve degenerates it can become a genus-five
curve of generic kind, and $F^{(5)}$ does {\it not} vanish
identically even on the moduli space.

\section{OPSMY ansatz}
\label{s:opsmy}

Another ansatz for superstring measures was proposed by
M.Oura, C.Poor, R.Salvati Manni and D.Yuen in their paper \cite{OPSMY}.
It is expressed in terms of \textit{lattice theta constants of
16-dimensional unimodular lattices}, well familiar from the
study of string compactifications  \cite{MorOlsh}.
Still we need to remind what they are.

An $h$-dimensional lattice $\Lambda$ is a subset of $\RF^h$
which is spanned by linear combinations with integer coefficients
of some $h$ linearly independent vectors.
These $h$ vectors together are called the \textit{basis} of the lattice.

Naturally associated with the lattice $\Lambda$ is a
genus-$g$ lattice theta constant:
a function on the Siegel space $\Hs$, defined as
\eq{
\ths_{\Lambda}\br{\tau} \defeq \su{\br{\vec p_1,\dots,\vec p_g} \in
\Lambda^g}\exp\big(\pi i (\vec p_k\cdot \vec p_l)\tau_{kl}\big),
}
where $(\vec p \cdot \vec p')$ denotes the usual Euclidean scalar product
of vectors in $\RF^h$, and summation is always assumed
over repeated
indices $k$ and $l$.
Lattice theta-constants have a very simple factorization
property: they remain themselves:
\be
\ths_{\Lambda}^{(g_1+g_2)}\!\!
\left(\begin{array}{cc} \tau^{(g_1)} & 0 \\
0 & \tau^{g_2)}\end{array}\right) =
\ths_{\Lambda}^{(g_1)}\br{\tau^{(g_1)}}
\ths_{\Lambda}^{(g_2)}\br{\tau^{(g_2)}}
\label{factlat}
\ee

A lattice is called \textit{self-dual}, or \textit{unimodular},
if it coincides with its dual lattice, i.e. if $\Lambda = \Lambda^*$.
The dual lattice $\Lambda^*$ is defined as the set of all vectors
$\vec u$ of $\RF^h$ such that $(\vec u\cdot \vec v)\in \ZG$ \textit{for all}
$\vec v \in \Lambda$.
A lattice is called \textit{even} if Euclidean norms of all
basis vectors are even. Otherwise the lattice is called \textit{odd}.

Lattice theta constant corresponding to self-dual $h$-dimensional
lattice with $h$ divisible by $8$ is a modular form of weight $h/2$
w.r.t. $\Gamma_g\br{1,2}$ if the lattice is odd,
and w.r.t. the full $\Gamma_g$ when the lattice is even.
Therefore lattice theta constants of 16-dimensional self-dual lattices are
semi-modular forms of the weight $8$ and
of particular interest for building superstring measures.

There are exactly eight 16-dimensional self-dual lattices \cite{Sloane},
all of them can be obtained from root lattices of some Lie algebras.
We list them with convenient notations in the following table:

\centerline{
\begin{tabular}{|c|c|c|}
\hline
Short  notation for lattice theta constant & Lattice & Gluing vectors \\
\hline
\hline
$\ths_0$&$\ZG^{16}$ & -- \\
\hline
$\ths_1$&$\ZG^8\oplus E_{8}$ & -- \\
\hline
$\ths_2$&$\ZG^4\oplus D_{12}^+$ & $\br{0^4,\frac{1}{2}^{12}}$ \\
\hline
$\ths_3$&$\ZG^2\oplus \br{E_7\oplus E_7}^+$ &
$\br{\frac{1}{4}^6,-\frac{3}{4}^2,\frac{1}{4}^6,-\frac{3}{4}^2}$ \\
\hline
$\ths_4$&$\ZG\oplus A_{15}^+$ & $\br{\frac{1}{4}^{12},-\frac{3}{4}^4},
\br{\frac{1}{2}^8,-\frac{1}{2}^8}, \br{\frac{3}{4}^4,-\frac{1}{4}^{12}}$\\
\hline
$\ths_5$&$\br{D_8\oplus D_8}^+$ & $\br{\frac{1}{2}^8,0^7,1}$ \\ \hline
\hline
$\ths_6$&$E_8 \oplus E_8$ & -- \\
\hline
$\ths_7$&$D_{16}^+$ & $\br{\frac{1}{2}^{16}}$ \\
\hline
\end{tabular}
}

\bigskip

The first column lists the naming conventions for theta constants
corresponding to lattices, the second column presents lattices themselves
and the third column contains \textit{gluing vectors} of particular lattices.
Here $\ZG^n$ denotes the trivial integer lattice $\ZG^n \subset \RF^n$.
$A_k$, $D_k$ and $E_k$ stand for root lattices of the
corresponding Lie algebras.
Cross over the name of a lattice (i.e. $\Lambda^+$)
means the union of this lattice
with lattices obtained by shifting it by all of its gluing vectors.
For notation to be concise, we follow \cite{Sloane}
and write $a^n$ for $a,\dots,a$ with n instances of $a$
(e.g. $\br{\frac{1}{2}^2,-1^2,0}$ means
$\br{\frac{1}{2},\frac{1}{2},-1,-1,0}$).
Also note that $E_8 = D_8^+$ with the gluing vector $\br{\frac{1}{2}^8}$.
First six lattices in the table are odd and the last two are even.

Note that in the paper \cite{OPSMY} the short notations
$\ths_i$ are also used for the lattice theta constants, but our convention
for numbering them is \textit{different}.
We have a reason for that, which will be explained at the end of
Section \ref{s:ltcvsotc}.

OPSMY use these lattice theta constants to write ansatz
for superstring measures \cite{OPSMY}.
Their idea is as follows: because all these lattice theta constants
are modular forms of weight 8 with respect to subgroup $\Gamma\br{1,2}$
of the modular group,
it is natural to try to build $\Xi[0]$ out of them and then obtain
all $\Xi[e]$ by acting on $\Xi[0]$ by modular transformations.

In fact OPSMY prove that up to genus 4 these lattice theta constants
span the \textit{entire} space of modular forms of weight 8
with respect to $\Gamma(1,2)$ and also find all the linear relations
among them:

$g=1:$
\eq{\ths_2=\frac{3}{2}\ths_1 - \frac{1}{2}\ths_0 \nn}
\eq{\ths_3=\frac{7}{4}\ths_1 - \frac{3}{4}\ths_0 \nn}
\eq{\ths_4=\frac{15}{8}\ths_1 - \frac{7}{8}\ths_0 \nn}
\eq{\ths_5=2\ths_1-\ths_0 \label{rel1}}

$g=2:$
\eq{\ths_3=\frac{7}{4}\ths_2-\frac{7}{8}\ths_1+\frac{1}{8}\ths_0\nn}
\eq{\ths_4=\frac{35}{16}\ths_2-\frac{45}{32}\ths_1+\frac{7}{32}\ths_0\nn}
\eq{\ths_5=\frac{8}{3}\ths_2-2\ths_1+\frac{1}{3}\ths_0}

$g=3:$
\eq{\ths_4 = \frac{15}{8}\ths_3-\frac{35}{32}\ths_2
+\frac{15}{64}\ths_1-\frac{1}{64}\ths_0\nn}
\eq{\ths_5 = \frac{64}{21}\ths_3-\frac{8}{3}\ths_2
+\frac{2}{3}\ths_1-\frac{1}{21}\ths_0}

$g=4:$
\eq{\ths_5 \mathop{=}_{\M} \frac{1024}{315}\ths_4
- \frac{64}{21}\ths_3+\frac{8}{9}\ths_2-\frac{2}{21}\ths_1
+\frac{1}{315}\ths_0 \label{rel4}}
Here $\M$ under the equation sign means that the equality is valid
when restricted to Jacobian locus.

Actually, in the last line a combination $\ths_6-\ths_7$ also enters,
but it vanishes on Jacobian locus.
So the linear relation looks as above only on Jacobian locus.
In fact, the combination $\ths_6-\ths_7$ would be proportional
to the cosmological constant. For $g\geq 5$ all 8 lattice theta constants
(including even ones) are linearly independent,
and it is unknown if they span the entire space of
$\Gamma(1,2)$-modular forms.

Let us now describe the OPSMY ansatz for superstring measures.
Our formulas will look a little different from \cite{OPSMY}.
First, we choose different enumeration of lattice theta constants,
as it was already mentioned, and, second, we choose different
normalization for the $\Xi$ function for genus 1, so that
\eq{
\Xi^{(1)}[0] = \ths_0-\ths_1 =
\theta^{16}[0]-\frac{1}{2}\theta^{8}[0]\su{e}\theta^8[e]
}
instead of
\eq{
\Xi^{(1)}_{OPSMY}[0] = \frac{1}{16}\ths_0-\frac{1}{16}\ths_1 =
\frac{1}{16}\theta^{16}[0]-\frac{1}{32}\theta^{8}[0]\su{e}\theta^8[e]
}
which is chosen in \cite{OPSMY}.
Therefore $\Xi^{(g)}$ would differ by a $2^{4g}$ factor.

Let $M$ be the following $6\times 6$ matrix with indexes running from 0 to 5:
\eqs{
&M_{ij} = 2^{j(1-i)}, \quad i=0..4,\; j=1..5, \nn\\
&M_{i0} = 1, \qquad i=0..5,\nn\\
&M_{5j} = 0, \qquad j=1..5
}
Explicitly,
\eq{
M = \left( \begin {array}{cccccc} 1&2&4&8&16&32\\
\noalign{\medskip}1&1&1&1&1&1\\\noalign{\medskip}1&\frac{1}{2}&
\frac{1}{4}&\frac{1}{8}&\frac{1}{16}&\frac{1}{32}\\
\noalign{\medskip}1&\frac{1}{4}&\frac{1}{16}&{\frac {1}{64}}&
{\frac {1}{256}}&{\frac {1}{1024}}\\
\noalign{\medskip}1&\frac{1}{8}&{\frac {1}{64}}&
{\frac {1}{512}}&{\frac {1}{4096}}&{\frac {1}{32768}}\\
\noalign{\medskip}1&0&0&0&0&0\end {array} \right)
}
Then $\Xi[0]$ functions of OPSMY ansatz in genera $g \leq 5$ are
\eq{
\Xi^{(g)}[0] = \summ{k=0}{5}\br{M^{-1}}_{gk}\ths^{(g)}_k, \quad
g = 1\ldots 5
}
(for genus 5 OPSMY, actually, propose different expression, see below).
In this ansatz we cannot write expressions of $\Xi[e]$ for general
$e$ as explicitly as in Grushevsky ansatz,
$\Xi[e]$ are obtained from $\Xi[0]$ by action of particular element
$\gamma_e$ of the modular group.

Explicitly,
\eq
{
\Xi^{(1)}[0]=\frac{1}{630}\,\ths_{{0}}-\frac{2}{21}\,\ths_{{1}}
+{\frac {16}{9}}\,\ths_{{2}} -{\frac {256}{21}}\,\ths_{{3}}
+{\frac {8192}{315}}\,\ths_{{4}} -{\frac {31}{2}}\,\ths_{{5}},
\nn}
\eq
{
\Xi^{(2)}[0]=-\frac{1}{42}\,\ths_{{0}} +{\frac {29}{21}}\,
\ths_{{1}}-24\,\ths_{{2}} +{\frac {2944}{21}}\,\ths_{{3}}
-{\frac {4096}{21}}\,\ths_{{4}} +{\frac {155}{2}}\,\ths_{{5}},
\nn}
\eq
{
\Xi^{(3)}[0]=\frac{1}{9}\,\ths_{{0}} -6\,\ths_{{1}}
+{\frac {808}{9}}\,\ths_{{2}} -384\,\ths_{{3}}
+{\frac {4096}{9}}\,\ths_{{4}} -155\,\ths_{{5}},
\nn}
\eq
{
\Xi^{(4)}[0]=-{\frac {4}{21}}\,\ths_{{0}}
+{\frac {184}{21}}\,\ths_{{1}} -96\,\ths_{{2}}
+{\frac {7424}{21}}\,\ths_{{3}}
-{\frac {8192}{21}}\,\ths_{{4}} +124\,\ths_{{5}},
\nn}
\eq
{
\Xi^{(5)}[0]={\frac {32}{315}}\,\ths_{{0}}
-{\frac {64}{21}}\,\ths_{{1}} +{\frac {256}{9}}\,\ths_{{2}}
-{\frac {2048}{21}}\,\ths_{{3}}
+{\frac {32768}{315}}\,\ths_{{4}} -32\,\ths_{{5}}
}
Substituting linear relations from (\ref{rel1})-(\ref{rel4}), we obtain:
\eql{oua1}{
\Xi^{(1)}[0]=\ths_{{0}}-\ths_{{1}},
}
\eq{
\Xi^{(2)}[0]=\frac{2}{3}\,\ths_{{0}}-2\,\ths_{{1}}+\frac{4}{3}\,\ths_{{2}},
}
\eq{
\Xi^{(3)}[0]=\frac{8}{21}\,\ths_{{0}}-\frac{8}{3}\,\ths_{{1}}
+\frac{16}{3}\,\ths_{{2}}-{\frac {64}{21}}\,\ths_{{3}},
}
\eql{oua4}{
\Xi^{(4)}[0]={\frac {64}{315}}\,\ths_{{0}}-{\frac {64}{21}}\,\ths_{{1}}
+{\frac {128}{9}}\,\ths_{{2}}-{\frac {512}{21}}\,\ths_{{3}}
+{\frac {4096}{315}}\,\ths_{{4}},
}
\eql{oua5}{
\Xi^{(5)}[0]={\frac {32}{315}}\,\ths_{{0}}-{\frac {64}{21}}\,\ths_{{1}}
+{\frac {256}{9}}\,\ths_{{2}} - {\frac {2048}{21}}\,\ths_{{3}}
+{\frac {32768}{315}}\,\ths_{{4}}-32\,\ths_{{5}}
}

It turns out that for genus 5 the cosmological constant
for this $\Xi^{(5)}$ is still proportional to $\ths_6-\ths_7$
and therefore is non-vanishing \cite{GruSM}.
Therefore OPSMY propose to cure it in the same way which was used with
the Grushevsky ansatz.
They find the numerical value of ratio of $\su{e}\Xi[e]$ and
$\ths_6-\ths_7$ (strictly speaking, they obtain the value of
this ratio only on Torelli space, which is a covering of moduli space,
not on Siegel half-space) and subtract from $\Xi[0]$ this expression
with this coefficient, divided by the number of characteristics.
In our normalization this looks like
\eq{
\widetilde{\Xi}^{(5)}[0]={\frac {32}{315}}\,\ths_{{0}}
-{\frac {64}{21}}\,\ths_{{1}}+{\frac {256}{9}}\,\ths_{{2}}
- {\frac {2048}{21}}\,\ths_{{3}}+{\frac {32768}{315}}\,\ths_{{4}}
-32\,\ths_{{5}}-\frac{686902}{24255}\ths_6 + \frac{686902}{24255}\ths_7
}

This ansatz cannot be continued to $g=6$ and above.
One can easily understand the problem, for example,
in the following way.
The factorization property requires that,
\eq{
\Xi^{(6)}\br{\tau^{(6)}_{1+1+1+1+1+1}}
= \Xi^{(1)}\br{\tau^{(1)}_1}\,\Xi^{(1)}\br{\tau^{(1)}_2}\,
\Xi^{(1)}\br{\tau^{(1)}_3}\,\Xi^{(1)}\br{\tau^{(1)}_4}\,
\Xi^{(1)}\br{\tau^{(1)}_5}\,\Xi^{(1)}\br{\tau^{(1)}_6},
}
and if we had $\Xi^{(6)}$ expressed as a linear combination of
lattice theta constants
$\Xi^{(6)} = \summ{p=0}{7}\al_p\ths_p$, then, using (\ref{factlat}),
\eqm{
\summ{p=0}{7}\al_p\ths_p^{(1),1}\,\ths_p^{(1),2}\,\ths_p^{(1),3}\,
\ths_p^{(1),4}\,\ths_p^{(1),5}\,\ths_p^{(1),6}= \\
=\br{\ths^{(1),1}_{{0}}-\ths^{(1),1}_{{1}}}
\br{\ths^{(1),2}_{{0}}-\ths^{(1),2}_{{1}}}
\br{\ths^{(1),3}_{{0}}-\ths^{(1),3}_{{1}}}
\br{\ths^{(1),4}_{{0}}-\ths^{(1),4}_{{1}}}
\br{\ths^{(1),5}_{{0}}-\ths^{(1),5}_{{1}}}
\br{\ths^{(1),6}_{{0}}-\ths^{(1),6}_{{1}}}
}
For $g=1$ all theta constants can be
expressed in terms of three linearly independent
ones $\ths_0,\,\ths_1,\,\ths_6$ with the help of (\ref{rel1}). If one does this and
expands all brackets, one will obtain a system of linear
equations on $\bd{\al_i}$, because after passing to
linearly independent functions all coefficients before
monomials in them shall vanish if l.h.s. is subtracted from r.h.s. It turns out that in this $g=6$
case the system of equations simply does not have
solutions. Thus the factorization constraint cannot be satisfied.
For example, if one does the same things for $g=5$, one will obtain
result \re{oua5}.

\section{Lattice theta constants vs Riemann theta constants}
\label{s:ltcvsotc}

In this section we describe the relation between lattice and 
Riemann theta constants
and write down explicit formulae, expressing ones in terms of the others.
Namely, we prove that
\eql{lorel}{
\ths_p^{(g)} = 2^{-gp} \xi^{(g)}_p, \quad p=0 \dots 4
}
in any genus.
For $p=0$ and $p=1$ the statement is rather trivial and already known.

Consider the $p=2$ case.
First, note that the factor of $\theta_0^4$ is common to the l.h.s.
and to the r.h.s. of \re{lorel} in this case, so we divide it out.
Then for the right hand side we have
\eqm{
 \xi^{(g)}_2/\theta_0^4 = \sum_{e_1,e_2}^{N_e}
 \theta^4_{e_1}\theta^4_{e_2}\theta^4_{e_1+e_2} = \sum_{e_1,e_2}^{N_e}\;
 \mathop{\mathop{\su{\vec n^a_I \in \ZG^g,}}_{\scriptscriptstyle a = 1..4,}
 }_{\scriptscriptstyle I \in 2^{\bd{1,2}}\setminus\br{\emptyset}}
 \exp \Bigg(\pi i \Bigg(\su{a}\br{\vec n^a_1+\frac{\vec \de_1}{2}}^T
 \tau\br{\vec n^a_1+\frac{\vec \de_1}{2}}+\\+\su{a}\br{\vec n^a_2+
 \frac{\vec\de_2}{2}}^T
 \tau\br{\vec n^a_2+\frac{\vec\de_2}{2}}+\su{a}\br{\vec n^a_{12}+
 \frac{\vec\de_1+\vec\de_2}{2}}^T
 \tau\br{\vec n^a_{12}+\frac{\vec\de_1+\vec\de_2}{2}}+\\
 +\br{\su{a}\vec n^a_1}^T\vec\ep_1+\br{\su{a}\vec n^a_2}^T\vec\ep_2+
 \br{\su{a}\vec n^a_{12}}^T\br{\vec\ep_1+\vec\ep_2}\Bigg)\Bigg)
}
We can perform summation over $\vec\ep_1,\,\vec\ep_2$.
Since they take values in $\br{\ZG_2}^g$ and enter the expression
as factors of
\eq{
\exp\br{\pi i \br{\su{a} \vec n^a_{1}+\vec n^a_{12}}^T\vec\ep_1}
}
and
\eq{
\exp\br{\pi i \br{\su{a} \vec n^a_{2}+\vec n^a_{12}}^T\vec\ep_2}
}
then all terms, in which at least on element of integer vectors 
$\vec v_1=\sum_a\br{\vec n^a_{1}+\vec n^a_{12}}$ and 
$\vec v_2=\sum_a\br{\vec n^a_{2}+\vec n^a_{12}}$ is odd, vanish, 
and all terms with all these elements being even survive and acquire 
a factor of $2^{2g}$.
Therefore, expanding also the sum in $\vec\delta$, we obtain
\eq{
 \xi^{(g)}_2/\theta_0^4 = 2^{2g}\su{\br{\Lambda_2^+}^g}
 \exp\br{\pi i \summ{i,j =1}{g}\br{n_i,n_j}\tau_{ij}},
}
where $\Lambda_2 \subset \ZG^{12}$ is a 12-dimensional lattice defined as
\eq{
\Lambda_2 = \bd{\br{n^1_1,\dots,n^4_1,n^1_2,\dots,n^4_2,n^1_{12},
\dots,n^4_{12}}\in \ZG^{12} \bigg| \br{\su{a}n^a_1+\su{a}n^a_{12}}
\vdots 2,\;\br{ \su{a}n^a_2+\su{a}n^a_{12}}\vdots 2}
}
and
\eq{
\Lambda_2^+ = \Lambda_2 \cup \br{\Lambda_2+\vec\alpha} \cup 
\br{\Lambda_2+\vec\beta}
\cup \br{\Lambda_2+\vec\gamma},
}
where
\eqs{
\vec\alpha &=
\br{\frac{1}{2},\frac{1}{2},\frac{1}{2},\frac{1}{2},\
0,0,0,0,\ \frac{1}{2},\frac{1}{2},\frac{1}{2},\frac{1}{2}}, \\
\vec\beta &=
\br{0,0,0,0,\ \frac{1}{2},\frac{1}{2},\frac{1}{2},\frac{1}{2},\
\frac{1}{2},\frac{1}{2},\frac{1}{2},\frac{1}{2}},\\
\vec\gamma &=
\br{\frac{1}{2},\frac{1}{2},\frac{1}{2},\frac{1}{2},\
\frac{1}{2},\frac{1}{2},\frac{1}{2},\frac{1}{2},\ 0,0,0,0}
}

Therefore, since factors $2^{2g}$ and $2^{-2g}$ perfectly cancel 
each other, to prove the lemma we only need to prove
that $\Lambda_2^+$ turns to the usual representation of $D_{12}^+$
under some orthogonal transformation
$A: \RF^{12} \rightarrow \RF^{12}$.
We now show that this transformation is the following one:
\eq{
A_{12} = \frac{1}{2}\br{
\begin {array}{ccc}
H_4&0&0\\
0&H_4&0\\
0&0&H_4
\end {array}
},
}
where $H_4$ is the following matrix
\eq{
H_4 = \br{
\begin {array}{cccc}
1&1&1&1\\
1&-1&1&-1\\
1&1&-1&-1\\
1&-1&-1&1
\end {array}
}
}
$H_4$ is a so-called Hadamard matrix, see below for details.

The $D_{12}^+$ lattice is defined as
\eq{
D_{12}^+ = \bd{\br{m_1,\dots,m_{12}}\in
\ZG^{12}\cup\br{\ZG^{12}+\br{\frac{1}{2}^{12}}}
\bigg| \br{\su{k}m_k}\vdots\, 2}
}
It is then straightforward to check that $A$ maps
every point of $\Lambda_2^+$
into point of $D^+_{12}$ and vice versa.
Indeed, $\Lambda_2$ consists of vectors $(e,e,e)$ and $(o,o,o)$,
where $e$ is either $(0,0,0,0)$ or $(1,1,1,1)$ or
one of the $C_2^4=6$ vectors like $(1,1,0,0)$,
while $o$ is either of the type $(1,0,0,0)$ or $(1,1,1,0)$.
All entries are defined modulo $2$.
So,
\be
\Lambda_2 = \Big\{(e,e,e),(o,o,o)\Big\}
\ee
Similarly
\be
D_{12} = \Big\{(e,e,e),(e,o,o),(o,e,o),(o,o,e)\Big\}
\ee
Now, denote by check a 4-vector shifted by $\br{1/2,1/2,1/2,1/2}$.
Then
$$\Lambda_2^+ = \Big\{
(e,e,e), (\check e,\check e,e), (\check e,e,\check e),
(e,\check e,\check e),
(o,o,o), (\check o,\check o,o), (\check o,o,\check o),
(o,\check o,\check o)\Big\},$$
while
$$D^+_{12} = \Big\{
(e,e,e), (\check e,\check e,\check e), (e,o,o),
(\check e,\check o,\check o),
(o,e,o), (\check o,\check e,\check o), (o,o,e),
(\check o,\check o,\check e)\Big\}$$

The action of $\frac{1}{2}H_4$, $\br{\frac{1}{2}H_4}^2=I$ is as follows:
$$
\begin{array}{ccc}
(0,0,0,0) & \leftrightarrow & (0,0,0,0)\\
(1,1,0,0) & \leftrightarrow & (1,0,1,0)\\
(1,1,1,1) & \leftrightarrow & (2,0,0,0)\\
&&\\
(1,0,0,0) & \leftrightarrow & \br{1/2,1/2,1/2,1/2}\\
(1,1,1,0) & \leftrightarrow & \br{3/2,1/2,1/2,-1/2}\\
&&\\
\br{-1/2,1/2,1/2,1/2} & \leftrightarrow &  \br{1/2,-1/2,-1/2,-1/2}
\end{array}
$$
what implies that $e\leftrightarrow e$,
$o \leftrightarrow \check e$ and
$\check o \leftrightarrow \check o$,
so that, for $A_{12}$,
$$
\begin{array}{ccc}
(e,e,e) & \leftrightarrow &(e,e,e) \\
(o,o,o) & \leftrightarrow &(\check e,\check e,\check e)\\
(\check e,\check e,e)& \leftrightarrow &(o,o,e) \\
(\check o,\check o,o)& \leftrightarrow &(\check o,\check o,\check e)
\end{array}
$$
This proves that $A_{12}$ indeed converts $\Lambda_2^+$ into
$D^+_{12}$ and vice versa.

It is interesting that basically everything is done by rotating 
vectors with the help of an Hadamard matrix. An $h$-dimensional 
Hadamard matrix is a matrix formed by elements of $h$ $h$-dimensional 
vectors such that these elements can be only $1$ or $-1$ and all 
these $h$ vectors are mutually orthogonal. It is an open problem 
if there exists an Hadamard matrix in every dimension $4k$. For 
dimensions which are powers of $2$ there exists a so-called 
Sylvester construction of Hadamard matrices, which is built 
upon the following fact: if $h\times h$ matrix $H_h$ is an Hadamard 
matrix, then the following $2h \times 2h$ matrix will also be 
an Hadamard matrix:
\eq{
H_{2h}=\br{
\begin{array}{cc}
 H_h & H_h\\
 H_h & -H_h
\end{array}
}
}
Our $H_4$ matrix which was used to rotate lattice is of this 
very type: it is built by applying this construction two times 
to $1 \times 1$ matrix
\eq{\br{1}}

It turned out that for two other lattices, namely, with $p=3$ 
and $p=4$, everything can be done also with the help of an Hadamard 
matrix. However in these cases we need a $16\times16$ matrix 
$H_{16}$ rather then a $4\times 4$ Hadamard matrix $H_4$. 
$H_{16}$ is obtained from $H_4$ again by applying the above 
construction two times. We do not present here its explicit 
form, because it is too bulky, and it is straightforward to obtain it.

The proof for $p=3$ case is then as follows: after dividing out 
the $\theta[0]$ common part and performing summation over 
epsilons in a way analogous to $p=2$ case we obtain a sum 
over $g$ copies of some lattice $\Lambda_3^+$.  $\Lambda_3$ 
is a subset of integer lattice in 14-dimensional space with 
coordinates
\eq{
\br{n^1_1,n^1_2,n^1_{12},n^1_3,n^1_{13},n^1_{23},n^1_{123},
n^2_1,n^2_2,n^2_{12},n^2_3,n^2_{13},n^2_{23},n^2_{123}}.
}
Notation for indexes of $n$ is analogous to the one used in 
$p=2$ case.
Denote $n^1_1+n^2_1$ by $m_1$, $n^1_2+n^2_2$ by $m_2$ and so on. 
Then $\Lambda_3$ is defined by the following conditions:
\eql{l3c1}{
m_1+m_{12}+m_{13}+m_{123}\; \vdots\; 2,
}
\eql{l3c2}{
m_2+m_{12}+m_{23}+m_{123}\; \vdots\; 2,
}
\eql{l3c3}{
m_3+m_{13}+m_{23}+m_{123}\; \vdots\; 2
}
$\Lambda_3^+$ is then obtained from $\Lambda_3$ by shifting it 
by vectors
\eql{l3v1}{
\br{\frac{1}{2},0,\frac{1}{2},0,\frac{1}{2},0,\frac{1}{2},
\frac{1}{2},0,\frac{1}{2},0,\frac{1}{2},0,\frac{1}{2}},
}
\eql{l3v2}{
\br{0,\frac{1}{2},\frac{1}{2},0,0,\frac{1}{2},\frac{1}{2},0,
\frac{1}{2},\frac{1}{2},0,0,\frac{1}{2},\frac{1}{2}},
}
\eql{l3v3}{
\br{0,0,0,\frac{1}{2},\frac{1}{2},\frac{1}{2},\frac{1}{2},0,0
,0,\frac{1}{2},\frac{1}{2},\frac{1}{2},\frac{1}{2}}
}
and then taking union of three resulting lattices with the original 
one.

Then the final statement is that lattice $\br{E_7\oplus E_7}^+$
is turned into $\Lambda_3^+$ by rotation with orthogonal matrix
$A_{16}=\frac{1}{4}H_{16}$, where $H_{16}$ is the 16-dimensional
Sylvester-type Hadamard matrix.
A question may arise: how can we rotate {\it a priori} 14-dimensional
lattices into one another using a $16\times16$ matrix?
The answer is simple: the $E_7$ lattice is usually defined
in 8-dimensional space
(in fact it lies in a particular hyperplane inside 8-dimensional space),
and therefore $\br{E_7\oplus E_7}^+$ is naturally defined in
16-dimensional space,
not 14-dimensional one \cite{Sloane}.
Therefore we can simply rotate $\br{E_7\oplus E_7}^+$
with the help of $A_{16}$.
And then it turns out that the two auxiliary coordinates
(namely, the first and the ninth ones)
are zero for images of all vectors of $\br{E_7\oplus E_7}^+$!
Therefore we can just drop them out and obtain a 14-dimensional lattice.
Simple technical calculations, which we do not provide here,
show that this 14-dimensional lattice is indeed $\Lambda_3^+$
(basically these calculations consist of checking that image
of every basis vector can be shifted back by one of the three vectors
(\ref{e:l3v1}-\ref{e:l3v3})
to an integer vector satisfying conditions (\ref{e:l3c1}-\ref{e:l3c3})).
This finishes the proof of relation \re{lorel} for the $p=3$ case.

The proof for $p=4$ is absolutely analogous to $p=3$ one
except there will be 15-dimensional space instead of 14-dimensional.
It is interesting that everything is again achieved with the help
of the same $A_{16}$ matrix as in the previous case.

Now it is clear, why we chose our enumeration for lattice theta constants.
It was done because functions $\xi_p$ have natural enumeration,
and functions $\ths_p$ correspond to them for $p=0\ldots 4$.

We end this section by reminding the well-known formulae for
the lattice theta constants
for \textbf{even} 16-dimensional lattices,
i.e. for $E_8\oplus E_8$ and $D_{16}^+$:
\eq{
\ths^{(g)}_6 = 2^{-2g}\br{\su{e}\theta^8[e]}^2,
}
\eq{
\ths^{(g)}_7 = 2^{-g}\su{e}\theta^{16}[e]
}

\section{The strange lattice}
\label{s:sl}

In the previous section we discussed the fact that, for $p=0\ldots 4$,
functions $\ths_p$ coincide (up to a simple constant factor)
with the functions $\xi_p$.
That is, the lattice theta series for lattices
$\ZG^{16}$, $\ZG^{8}\oplus E_8$, $\ZG^{4}\oplus D_{12}^+$,
$\ZG^{2}\oplus \br{E_7\oplus E_7}^+$ and $\ZG\oplus A_{15}^+$
are equal to combinations of the same type of ordinary theta constants.

However, there is one more odd 16-dimensional lattice,
namely $\br{D_8\oplus D_8}^+$.
To our surprise, the corresponding lattice theta constant, $\ths_5$,
behaves very different from the others.

With the help of OPSMY linear relations (\ref{rel1})-(\ref{rel4})
between lattice theta constants we can express $\ths_5$ through
$\ths_0,\dots,\ths_4$ for every genus $g \leq 4$.
Then, knowing the formulae from the previous section,
expressing $\ths_0,\dots,\ths_4$ through the ordinary theta constants,
we can do the same with $\ths_5$ for all genera $g \leq 4$.
It turns out that these expressions in all these genera follow one
and the same pattern:
\eql{stl}{
\ths_5^{(g)} \mathop{=}_{\M} 2^{-\frac{g(g-1)}{2}}\,
\br{\prodd{i=1}{g}\br{2^i-1}}^{-1} \, G^{(g)}_g, \quad g \leq 4
}
Thus the last lattice theta constant surprisingly coincides
(again up to a constant factor, this time a little bit more sophisticated)
with an expression through the ordinary theta constants of the
\textit{second} type of the two mentioned in Section \ref{s:grush}.
This, however, is valid only on Jacobian locus,
because for genus 4, where for the first time Jacobian locus
differs from the  entire Siegel half-space,
the linear relation (\ref{rel4})
holds only on Jacobian locus.
On the entire Siegel half-space formula \re{stl} would then look like
\eqs{
\ths_5^{(4)} &= \frac{1}{2^{6}\cdot 315} \, G^{(4)}_4 + 
\frac{3}{7}\br{\ths_6-\ths_7} =\\
             &= \frac{1}{2^{6}\cdot 315} \, G^{(4)}_4 - 
             \frac{3}{2^{8}\cdot7}\,F^{(4)}
}

We could not prove the formula \re{stl}
by a direct method like the one used for other lattices.
This is unfortunate, because
starting from genus 5 there are no linear relations between
lattice theta constants.
Therefore a question arises: would lattice  theta constant
$\ths_5$ continue to follow the same pattern \re{stl}
for genera $g \geq 5$?
At the moment we cannot answer this question
due to various technical difficulties.
Of course, equality of type \re{stl} for genus 5 cannot
be valid on entire Siegel space, since the right hand side,
i.e. $G^{(5)}_5$, contains square roots of monomials in theta functions,
and therefore possesses singularities on the entire Siegel space.
However, R. Salvati Manni in the paper \cite{SM} argued that all
this singularities lie outside Jacobian locus.
Thus the equality could in principle be true,
if restricted on Jacobian locus, despite it would be a rather
mysterious theta-constant identity.
Perhaps, it can help to shed some light on the possible explicit form
of Schottky identities at higher genera.

\section{Relations between Grushevsky and OPSMY ans\"atze}
\label{s:arel}

In this section we briefly discuss relation between the
two ans\"atze for NSR measures.

If we substitute expressions \re{lorel} for the lattice theta constants
into the formulae (\ref{e:oua1}-\ref{e:oua4}) for OPSMY ansatz
for genera $g$ from 1 to 4, we straightforwardly obtain
formulae (\ref{e:gra1}-\ref{e:gra4}) for Grushevsky ansatz.
Therefore for $g \leq 4$ both ans\"atze coincide,
which is in perfect agreement with uniqueness properties
of OPSMY and Grushevsky ans\"atze.

For genus $g=5$ case it is difficult to compare these two ans\"atze
because we do not know how $\ths_5$ relates to ordinary theta constants.
We can do this if we assume that
\re{stl} continues to hold beyond $g\leq 4$,
i.e. that
\eq{
\ths_5^{(5)} \stackrel{?}{\mathop{=}_{\M}} 2^{-\frac{5(5-1)}{2}}\,
\br{\prodd{i=1}{5}\br{2^i-1}}^{-1} \, G^{(5)}_5
}
If we make this assumption, then we again can substitute all expressions
for lattice theta constants into expression \re{oua5}
for OPSMY ansatz and see that the resulting formula is the same that
\re{gra5} for Grushevsky ansatz.
Additional parts, entering modified expressions for ans\"atze
that come from subtracting cosmological constant divided by the number
of characteristics, are then also equal.
Thus, if the function $\ths_5$ continues to follow for $g=5$
the same pattern it followed for $g \leq 4$
-- despite $G^{(5)}_5$ contains square roots! --
then Grushevsky and OPSMY ans\"atze coincide at genus 5 too.

\section{Conclusion}

In this paper we discussed some relations between lattice and
Riemann theta constants
and ans\"atze for superstring measures which are written in terms of them.
We presented explicit formulae expressing lattice theta constants
for eight 16-dimensional self-dual lattices through Riemann
theta constants.
This was then used to explicitly show that Grushevsky
and OPSMY ans\"atze coincide for $g\leq 4$, as it was originally predicted.
However, already for $g=5$ there are difficulties
to see if the ans\"atze remain the same.
This is related to the strange behaviour of the theta constant,
associated with one of the lattices, $\br{D_8\oplus D_8}^+$.
It is a very interesting problem to see exactly, how it can be expressed
through Riemann theta constants for genera $g \geq 5$.
Coincidence of ans\"atze implies a very elegant
-- but also extremely surprising -- formula for
this lattice theta-constant, but we did not find an equally
nice way to prove (or to reject) it.

At the same time, despite the beauties,
once again found in this paper in the world of the modular forms,
it seems that hypothesis that NSR measures can be always expressed
in their terms is overoptimistic.
Most probably, like Mumford measure $d\mu$ itself,
the ratio of $d\mu[e]/d\mu$ is going to be a function on moduli
space without a natural non-singular continuation to entire
Siegel half-space. At best it can contain Schottky-related
forms in denominator, like it happens to $d\mu$ at genus $g=4$
\cite{bosM}.
To find these ratios for genera  $g\geq 5$ one should apply or invent
some other technique.

\section*{Acknowledgements}

We are grateful to Philipp Burda, Sam Grushevsky, Igor Krichever, 
Andrei Levin, Andrei Losev, Yuri Manin, Andrei Marshakov, Alexander 
Popolitov, Riccardo Salvati Manni, Jean-Pierre Serre, George Shabat,
Shamil Shakirov and Andrei Smirnov for illuminating discussions and remarks.
Our work is partly supported by Russian Federal Nuclear Energy
Agency, by the joint grants 09-02-91005-ANF, 09-02-90493-Ukr,
09-02-93105-CNRSL and 09-01-92440-CE,
by the Russian President's Grant of
Support for the Scientific Schools NSh-3035.2008.2, by RFBR grants 07-02-00645 (A.M., P.D-B.)
and 07-02-00878 (A.S.) and by Federal Agency for Science and Innovations of Russian Federation
(contract 02.740.11.5029) (P.D-B., A.S.).

\end{document}